\documentclass{IEEEtran}
\usepackage{graphicx}

\usepackage{caption}
                                                                                                              
\usepackage{fixltx2e}
\usepackage{url}

\newcommand{\shorttitle}[1]%
{\markboth{Proceedings of the 1\MakeLowercase{$^{st}$} ICST, Greece 2009}{#1} }
\newcommand{\etal}{\MakeLowercase{\textit{et al. }}} 
                                                                                                              
\begin{document}
\title{Fresnel Zone Plate Telescopes as high resolution imaging devices}


\author{\IEEEauthorblockN{Sandip K. Chakrabarti\IEEEauthorrefmark{1}\IEEEauthorrefmark{2}, 
Sourav Palit\IEEEauthorrefmark{2}, Anuj Nandi\IEEEauthorrefmark{2}\IEEEauthorrefmark{3}, 
Vipin  K.  Yadav\IEEEauthorrefmark{2}\IEEEauthorrefmark{3}, Dipak Debnath\IEEEauthorrefmark{2}}.\\
\IEEEauthorblockA{\IEEEauthorrefmark{1}S. N. Bose National Centre for Basic Sciences, Block-JD, 
Sector-III, Salt Lake, Kolkata - 700098, India. \\ Email: chakraba@bose.res.in}\\
\IEEEauthorblockA{\IEEEauthorrefmark{2} Indian Centre for Space Physics, 43-Chalantika, Garia Station
Road, Kolkata - 700084, India.\\ Email: sourav@csp.res.in, anuj@csp.res.in, vipin@csp.res.in, dipak@csp.res.in}\\
\IEEEauthorblockA{\IEEEauthorrefmark{3} Indian Space Research Organization - HQ, New BEL Road, Bangalore -
560231, India.}}

\shorttitle{Chakrabarti \etal FZP Telescopes}
\maketitle

\begin{abstract}

Combination of Fresnel Zone Plates (FZP) can make excellent telescopes for
imaging in X-rays. We present the results of our experiments with such telescopes
with an X-ray source kept at a distance of $45$ feet. We compare the patterns obtained
from experiments with those obtained by our Monte-Carlo simulations. In simulations,
we allow the sources to be at finite distances (diverging beam) as well as at infinite distances (parallel beam)
and show that the resolution is worsened when the source is nearby. We also present simulated 
results for the observation of the galactic center and show that the sources may be reconstructed with accuracy.
We compare the performance of such a telescope with other X-ray imaging devices used in space-astronomy.
The Zone Plate based instrument has been sent for the first time in a recently launched KORONAS-FOTON satellite.

\end{abstract}

\section{Introduction}
 
Zone Plate Telescopes (ZPTs) have generated immense theoretical interest
[1-7].
In Chakrabarti et al. [8] and Palit et al. [9], extensive theoretical
studies have been presented and some of the results of the experiments on such telescopes conducted at 
the Indian Centre for Space Physics X-ray laboratory. Such telescopes have already
been used in RT-2 Payloads aboard Russian satellite KORONAS-FOTON launched on 30th January, 2009.
The ZPTs have an advantage over other high resolution X-ray telescopes [10-11]
in that they can have arbitrarily high angular resolution and that the
resolution can also be independent of energy bands in a large range of energy. The only disadvantage
is that ZPTs are two-element systems as opposed to the
conventional coded aperture masks (CAMs) which are single element systems.
In a zone plate telescope, two plates are aligned but the source is kept off-axis in order to
get relative shift of the ray of light on the image plane. Sources at higher off-axis 
locations produce finer fringe separations. These fringes may or may not be separable 
depending on how fine the detector pixels are. The fringe patterns on the detector are 
inverse Fourier transformed to get the source location.

\section{Experimental Setup, Monte-Carlo simulations and the comparison of the Results}

Fig. 1a shows our experimental setup. It consists of an X-ray generator with a molybdenum target. The beam
line is 45 feet long. The anode voltage and the current can be controlled at will.
At the end of the beam-line, the tungsten zone plates (Fig. 1b) are placed which are followed by a CMOS
detector having the pixel size equal to 50 micron. The zone
plates we use are made up of $1$ mm thick tungsten, the `opaque' zones are opaque up to $\sim 100$keV.
We did not choose to vacuum the beam-line and supply high energy X-rays by
applying higher anode voltage instead. The intervening air column will cause an absorption in the soft X-rays but
hard X-rays including the K$_\alpha$ and K$_\beta$ lines will remain strong.

\begin{figure}[h]
\centering
\includegraphics[height=2.60in, width = 1.5in]{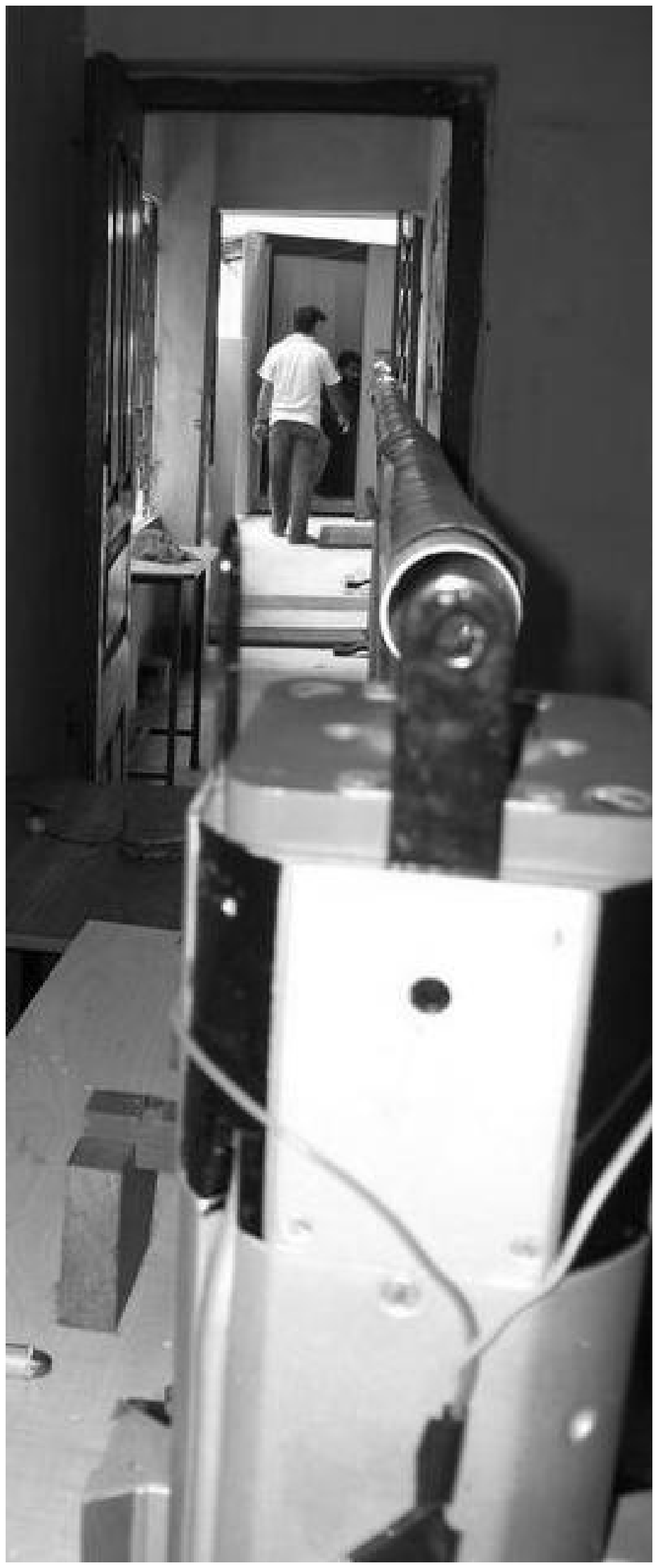} \vspace{0.15in}

\includegraphics[height=2.12in ]{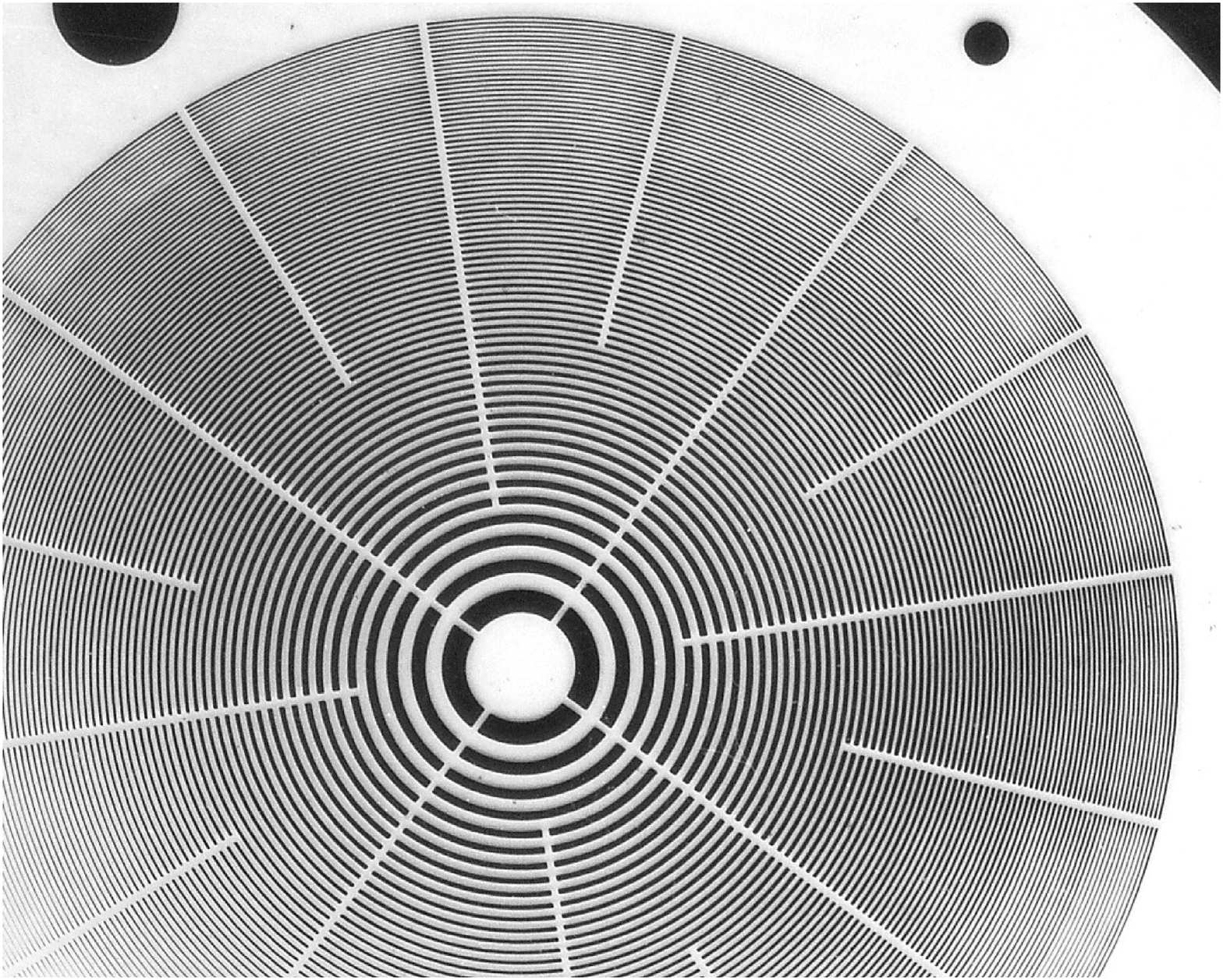}
\caption{(a) The 45 feet beam-line used in our experiments. At the near end is the
X-ray generator (0-50kV) with molybdenum target and at the far end is the detector assembly.
(b) A positive zone plate. [8]}
\end{figure}

\clearpage

Monte-Carlo simulations are done to reproduce circumstances when it is 
difficult to change the experimental set-up at will. We conduct simulations not only 
to reproduce the cases when experiments were carried out, but also by placing 
the detector at different distances, and with arbitrarily high or low photon fluxes. 
We also conduct simulations with multiple sources. The code was written in IDL
details are in [8-9].



Fig. 2a shows the shadow cast by two aligned zone plates when the source is almost
on the axis of the plate holder.  The circular fringes are of diameter
$r_{d,m} = r_1 [ \frac{z (2m-1)}{2D(1-\frac{D}{2z})}]^{1/2} $, where $r_1$ is the radius of the 
central zone, $z$ is the distance of the source from the first zone plate (facing the source),
$D$ is the distance between the zone plates and $m$ is the dark fringe number from the center.
In our case, $r_1=0.122$ cm, $D=20$ cm, $z=1338$ cm. In Fig. 2b, we show the results of the
simulation. Reconstructed image shows sharp rise of counts at the center.

\begin{figure}[h]

\centering{
\includegraphics[height=1.75in,width=1.85in]{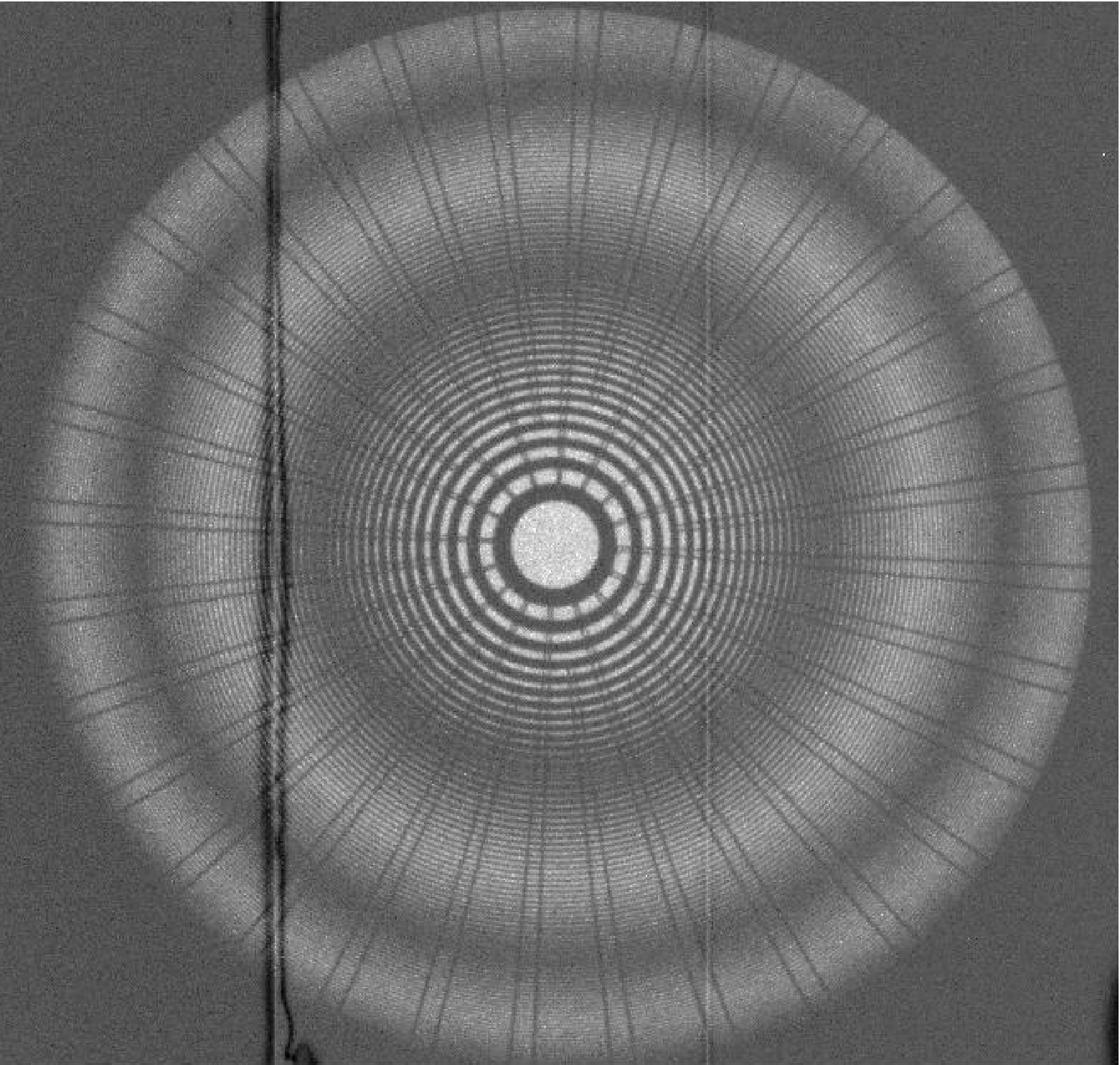} \vspace{0.08in}}

\hspace{0.62in}
\includegraphics[height=1.85in,width=2.60in]{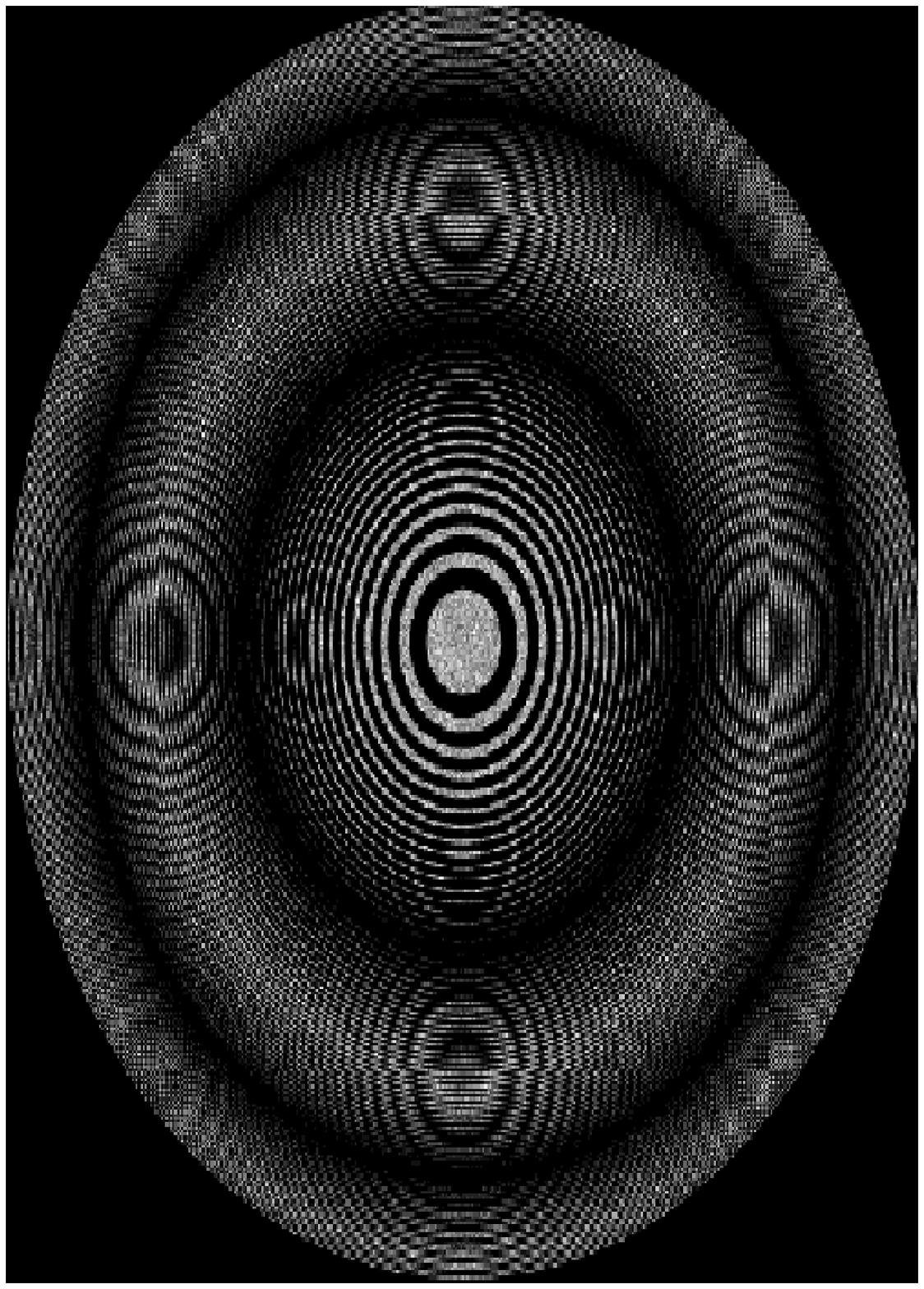}
\caption{Circular dark and bright circles obtained by an on-axis point source located
45 feet away. Two zone plates were used which are separated by $20$cm.
(a) Experimental result and (b) Monte-Carlo simulation results with $2\times 10^5$ photons [8].}
\label{}
\end{figure}

Figs. 3(a-b) show the fringes produced on the detector plane by a misaligned source: (a) is the 
actual experimental result on a CMOS detector and (b) is the simulation result. In Figs. 3(c-d),
the reconstructed images are shown. Since we used only one pair of zone plates, both the central
DC offset as well as the alias (mirror image) are seen. These would be removed by using four pairs 
of zone plates as shown below.

\begin{figure}[h]
\centering
\includegraphics[height=1.75in]{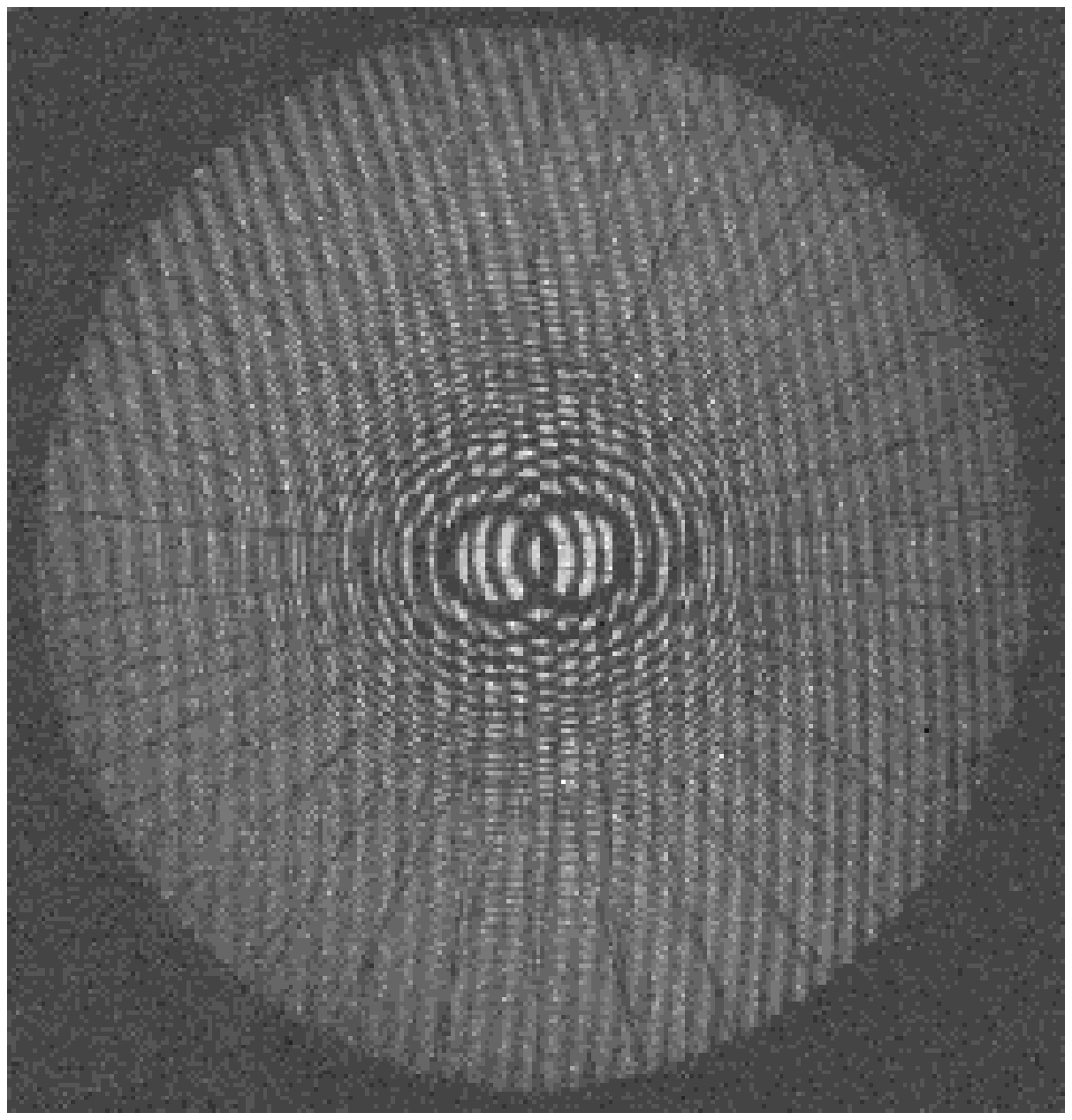} \vspace{0.1cm}
\includegraphics[height=1.75in]{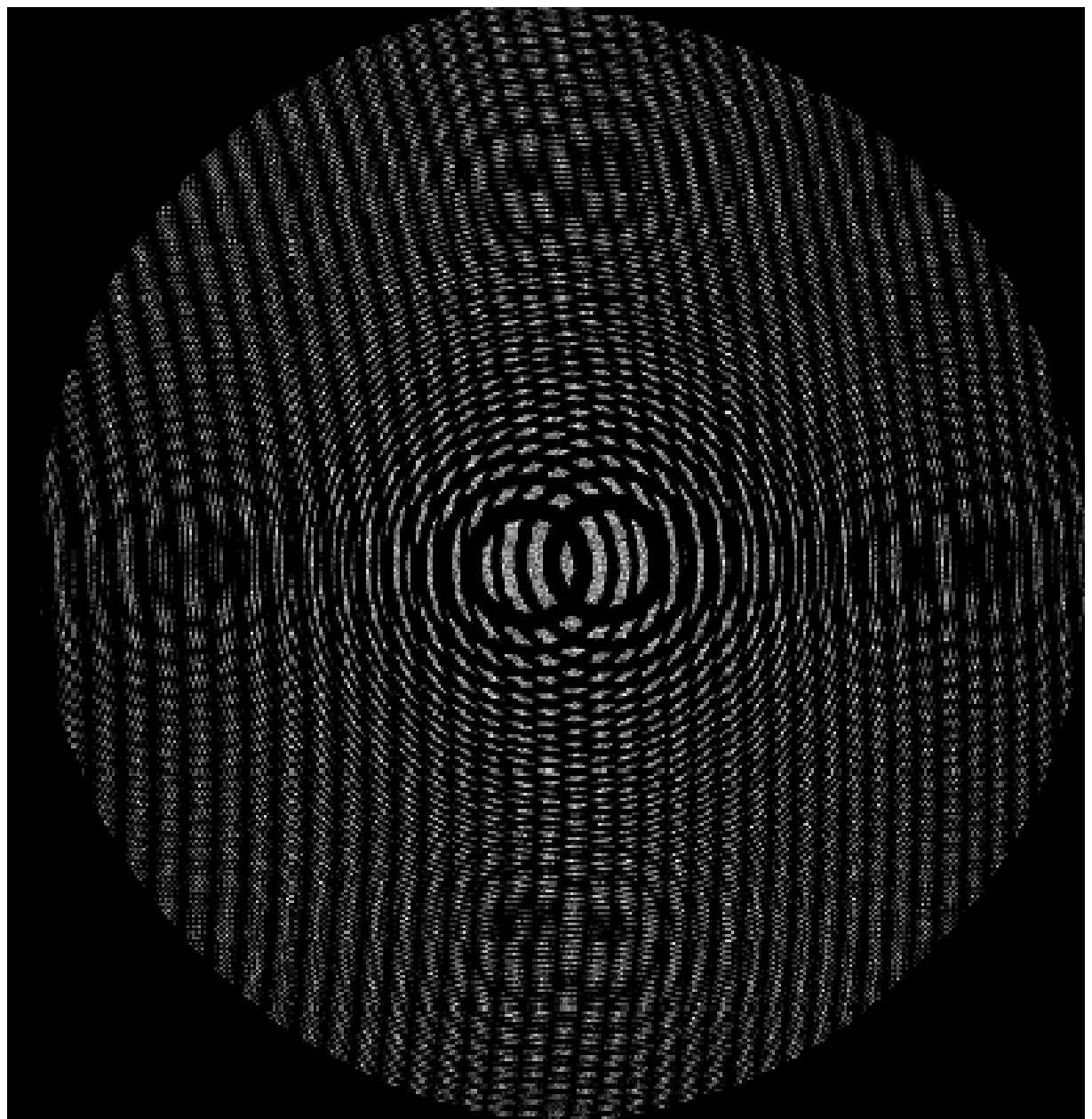}  \vspace{0.1cm}
\includegraphics[height=1.65in]{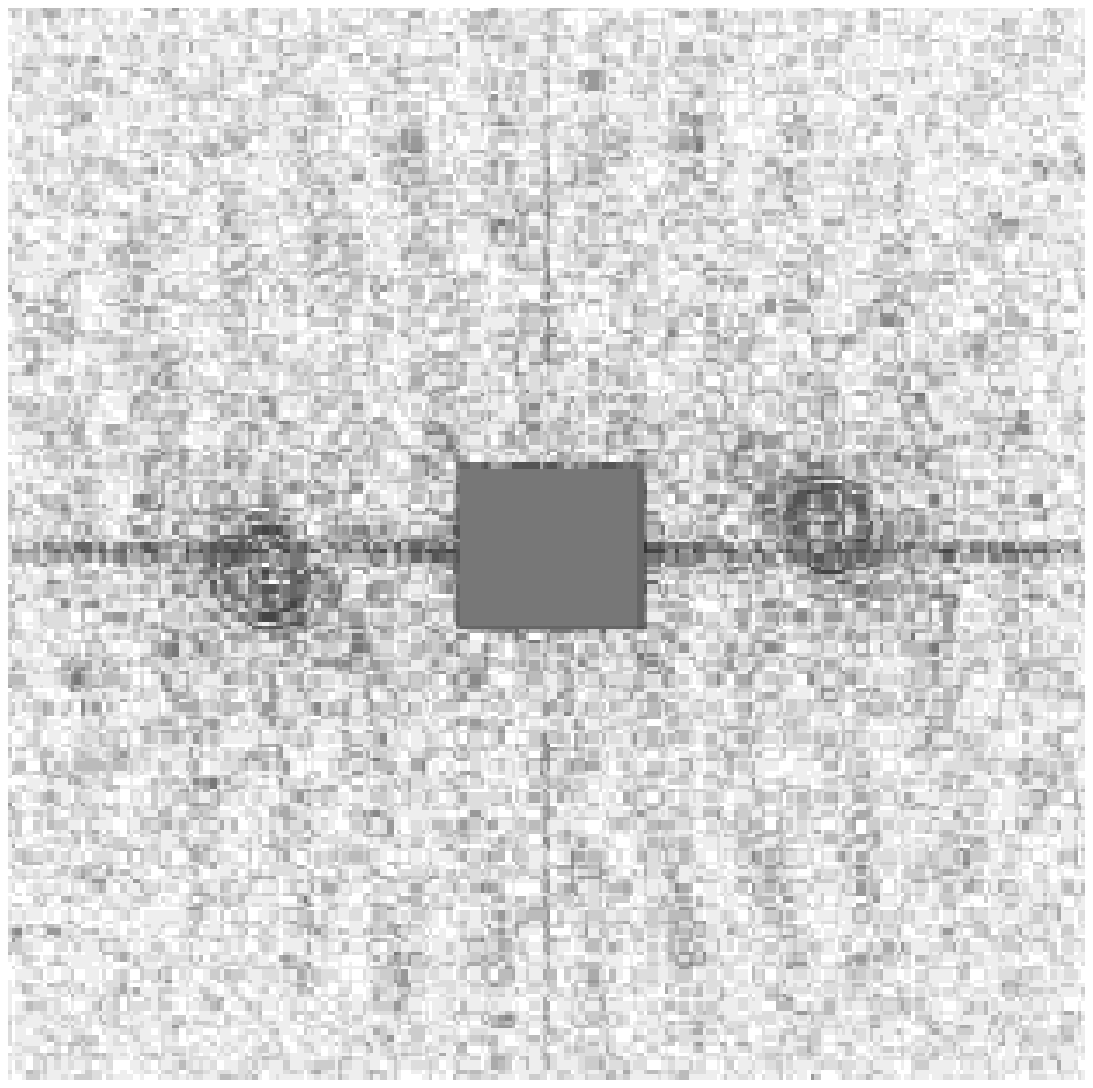} 
\includegraphics[height=1.65in]{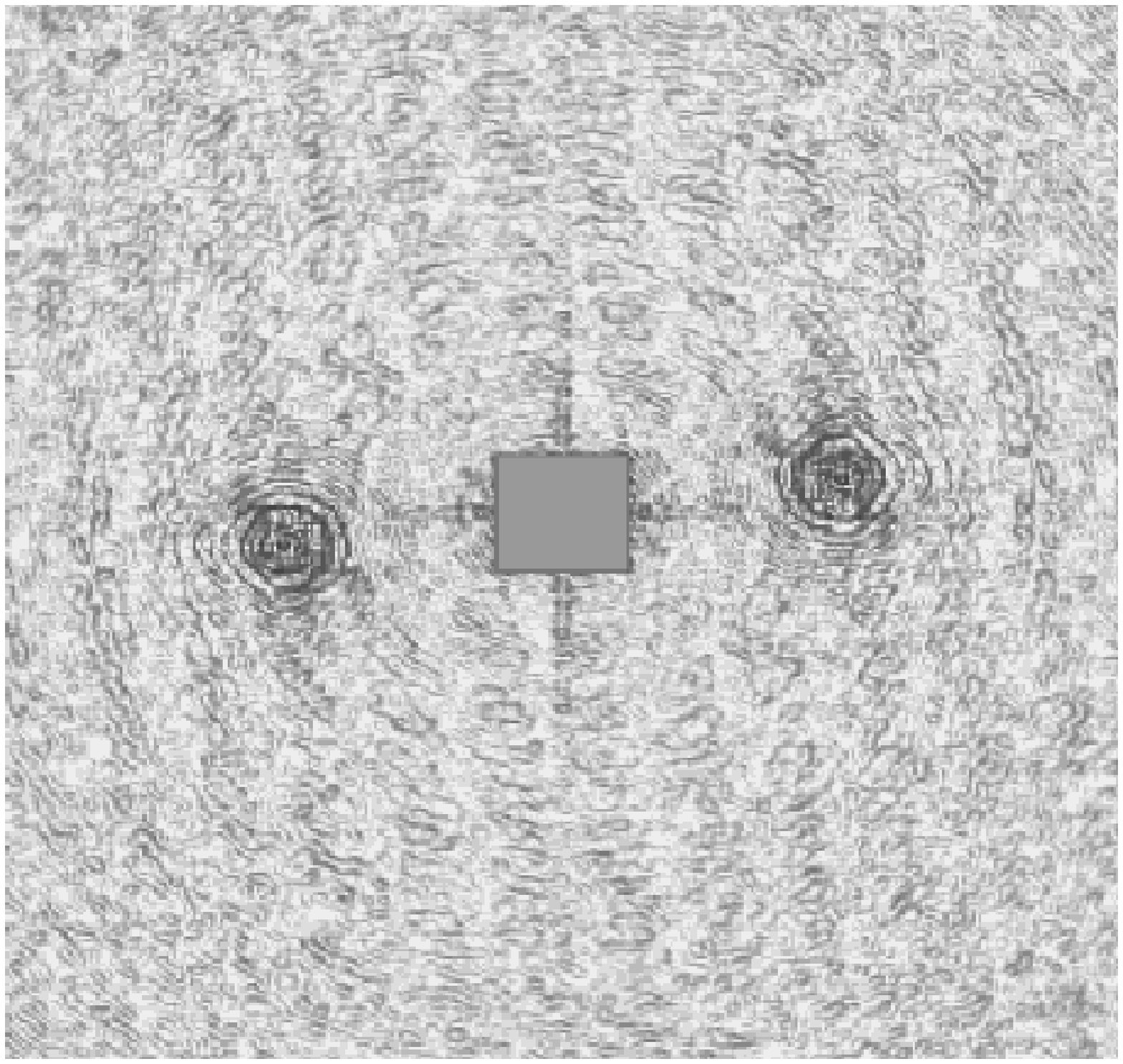}
\caption{Fringe patterns from the (a) experiment and the (b) simulation when the zone plates are
placed at a distance of $D=20$ cm and the source is $0.65$ degree away from the optical axis. $5 \times 10^6$
photons were used in the simulation. The deconvolved images from (a) and (b) are drawn in (c) and (d)
respectively. The image becomes sharper as the source is taken farther out [8].}
\end{figure}

First, we present the results of simulations with the source placed at infinity. The zone plate
spacing in each case is taken to be $10$ cm. The source is placed at an angular
distance of $\phi=1500$ arcsec from the optical axis and at a zenith angle
of $\theta=45$ deg measured from the positive X-axis. The number of photons infalling
on each of the front zone plates (ZP1s) is $10^5$. The fringes obtained in each of the pairs
is given in Fig. 4a. The source obtained by reconstruction is shown in Fig. 4b.
Neither the DC offset, nor the pseudo-source appears in the reconstructed image.

\begin{figure}[h]
\centering
\includegraphics[height=3.0in]{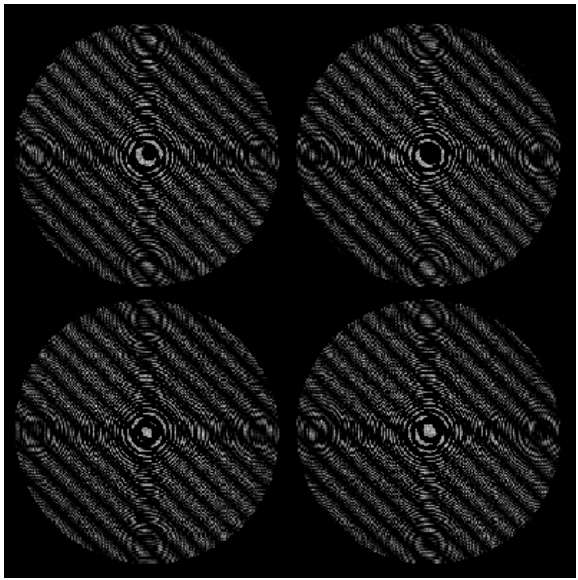}\vspace{0.5cm}
\includegraphics[height=2.5in,width=2.5in]{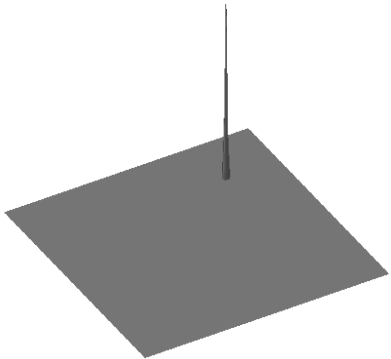}
\caption{(a) Fringes obtained for a source at an infinite distance with four pairs of zone plates.
(b) Reconstructed source on a CMOS detector ($4500$ arcsec along each side of the squared base.
Neither the pseudo-source nor the DC offset are seen in the reconstructed image [9].}
\end{figure}

When the source is at a finite distance, the DC off-set is always canceled but the pseudo-source cannot be 
canceled even if four pairs are used. This is because the angles subtended by the source
at different pairs are different.

For point sources at finite distances, when reconstructed from the fringe 
pattern, there are spreads. These are due to the point spread function, which 
assumes the shape of the aperture of imaging device(here the common area of the two zone plates 
intersected by the rays coming from the sources). This broadening worsens angular resolution for 
nearby sources.

To show this effect prominently we carry out the following simulation with two sources.
One is placed at $\phi = 2400$ arcsec and $\theta = 0$ degree and the other 
is placed at $\phi = 1200$ arcsec and $\theta = 90$ degree. 
The sources are placed at a distance of $45$ ft from ZP1. The zone plate separation is taken to be $100$ cm. 
In Fig. 5a, we show the fringe pattern and in Fig. 5b, we show the reconstructed image
($5800$ arcsec on each side). It is clear that locationwise, the sources have been placed properly, 
although they look `similar' to the  Moir\'{e} fringe patterns, which have special shapes due to 
off-axisness of the sources. In the simulations, the number of photons impinging from these 
two sources are taken to be $100000$ and $70000$ respectively, causing one pattern to be 
slightly brighter than the other. 

\begin{figure}[h]
\centering
\includegraphics[height=2.0in]{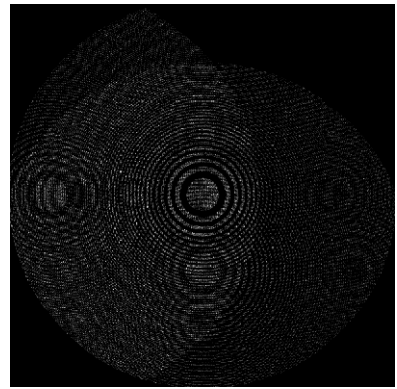}  


\vspace{0.2cm}
\includegraphics[height=2.0in]{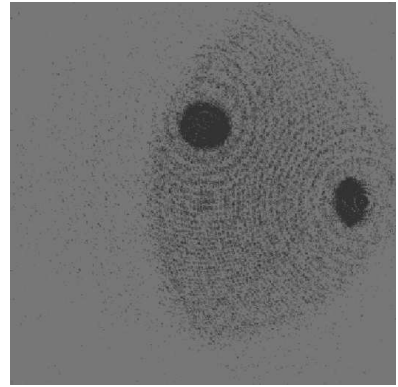}
\caption{(a) Moir\'{e} fringe patterns obtained for two sources placed at about $45$ ft away from the 
front zone plate. In the telescope, the plates are separated by $100$ cm. (b) The reconstructed 
images. Due to large off-axisness the fringes do not cover the entire zone plates. The image size
is compatible with the distance of the source and the image shape is compatible with the 
degree  off-axisness.}

\end{figure}

In order to demonstrate the capability of resolution of a ZPT, in Figs. 6(a-b), we show 
the reconstructed images of a pair of sources which were placed at an angular
distance of (a) $103$ arcsec and (b) $51.5$ arcsec (limiting case). Each side of the
base is $20$ arcmin in length. This agrees with the resolution of the instrument
$\theta=2\delta r/D$, where $\delta r$ is width of the finest ring of the zone plate and $D$
is the zone plate separation. In our case, $\delta r = 50$micron and $D=40$cm and hence the 
resolution is $51.5$ arcsec.

\begin{figure}[h]
\centering
\includegraphics[height=1.5in]{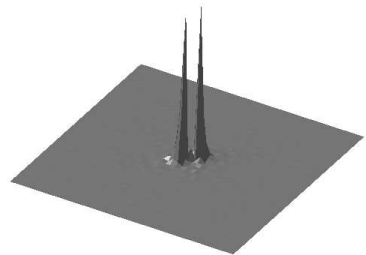} \hspace{0.5cm}
\includegraphics[height=1.5in]{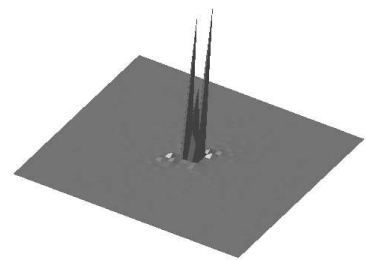}
\caption{Sources placed at infinity and separated by (a) $103$ arcsec and (b) $51.5$
arcsec respectively are reconstructed from the fringe patterns. The zone plate
separation is $40$ cm. In each Figure, the base is $20$ arcmin on each side [9].}
\label{}
\end{figure}

\clearpage
We now present the result of the simulation in which we examine how the fringe system should look like
when a prominent X-ray sources are turned on near the Galactic center. In Fig. 7a, we show the fringe patterns
produced by the four pairs of zone plates on a CMOS detector ($50$ micron pixel size).
The plate separation is $20$ cm. The reconstructed source distribution is seen in Fig. 7b.

\begin{figure}[h]
\centering{
\includegraphics[height=3.4in]{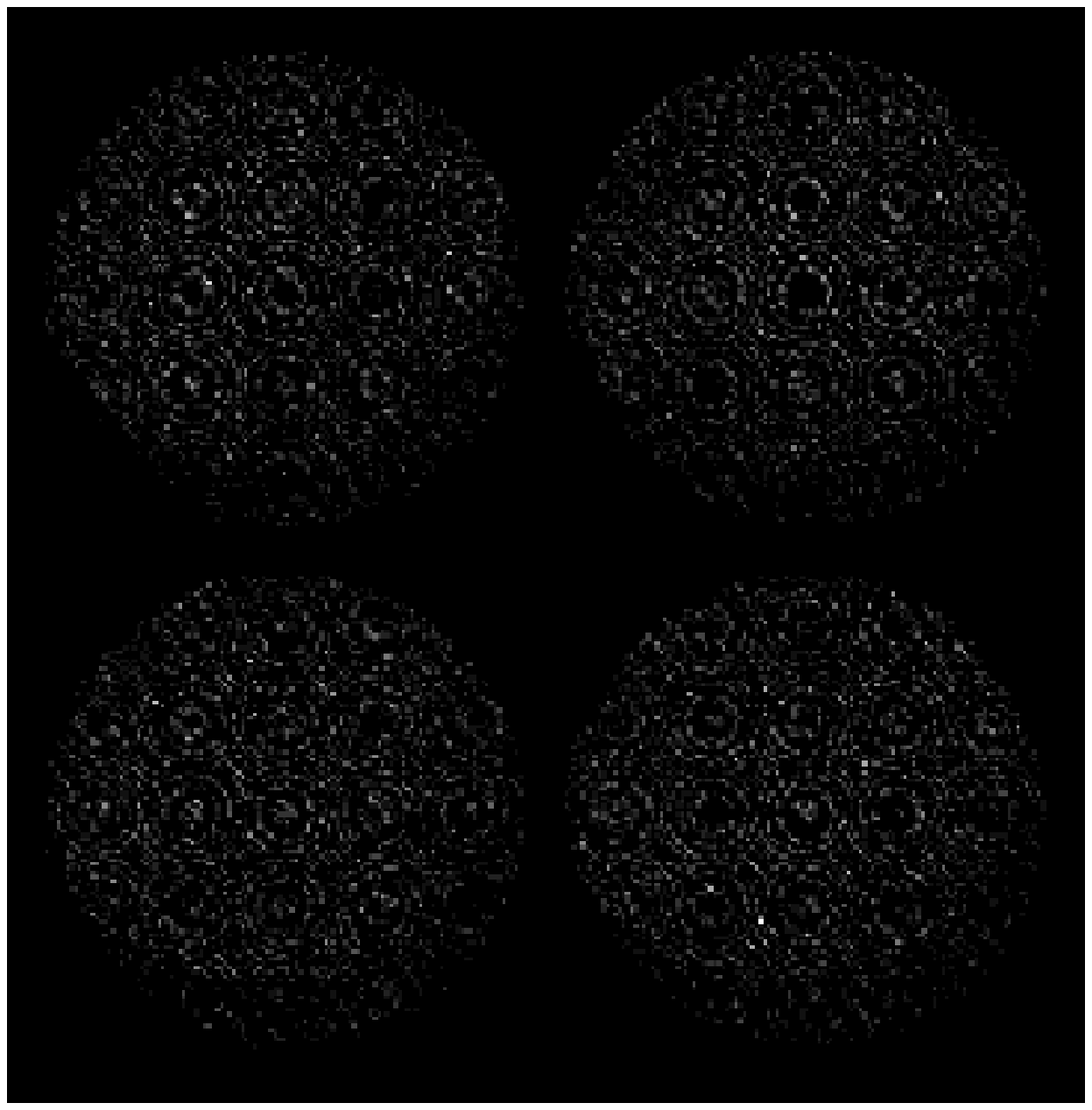}}


\includegraphics[height=3.0in]{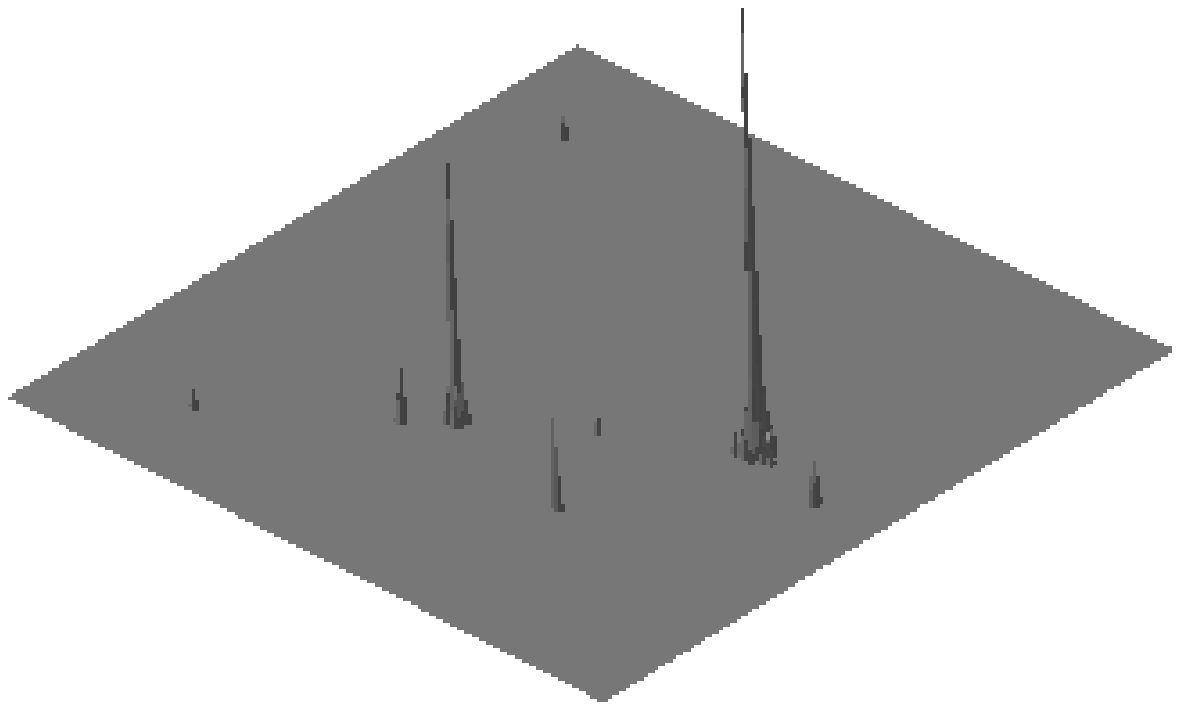}
\caption{(a) The fringe patterns with four pairs of zone plates for actual X-ray source
distributions near the Galactic center. (b) The reconstructed source distribution from the fringes.
The field of view is $3.5^o \times 3.5^o$ [9].}
\end{figure}

\section{Concluding remarks}

With the technological advancements it has become possible to produce very accurate large area 
zone plates having very fine rings using very heavy metals such as tungstens. We presented results
using pairs of tungsten zone plates which are of $\sim 3.0$cm diameter and show that achromatic
angular resolution of a few tens of arc seconds was possible even when the plate separation is 
only tens of centimeters. By increasing the separation of the zone plates, it is possible to obtain
arbitrarily high resolutions. With four pairs of zone plates (two for sine and two for cosine transforms),
very accurate imaging is possible. However, the ultimate result depends on the type of the detector
placed behind the telescope. In a CMOS detector, the information about energy is lost. In a CZT type
detector, the energy dependent imaging is possible, but the image resolution is compromised since the 
pixel size is 50 times bigger that that of CMOS.

Because the size of the zone plates having finer zones can be at the most a few cm, imaging faint 
sources would be impossible. Thus zone plates are most suited for transient bright and pointlike sources.
In the Russian solar mission named KORONAS-FOTON which was launched on the 30th January, 2009,
the zone plate set-up mentioned above is included in the Indian payload RT-2/CZT. The sun is 
still in the quiet phase and images can be obtained when it is more active.

\section*{Acknowledgment}

The expriments were possible due to a grant from ISRO to ICSP. The research of SP and DD at ICSP
is supported by CSIR.

\end{document}